\documentstyle[12pt]{article}

\def\'#1{{\accent19\ifx #1i \i\else #1\fi}}

\def\be{\begin{equation}}
\def\ee{\end{equation}}
\def\bea{\begin{eqnarray}}
\def\eea{\end{eqnarray}}

\newcommand{\boldmathgamma}{\mbox{\boldmath$\gamma$\unboldmath}}

\newcommand{\bfu}{{  u}}

\newcommand{\bfw}{{  w}}

\newbox\Ancha
\catcode`@=11
\newdimen\ex@
\title{Electroweakly  interacting scalar and gauge
bosons,  and leptons,   from field equations on spin 5+1
dimensional
 space}

\author{J. Besprosvany}

\date{Instituto de F\'{\i}sica, Universidad Nacional Aut\'onoma de M\'exico,
Apartado Postal 20-364, M\'exico 01000, D. F., M\'exico}

\begin{document}

\maketitle









\jot = 1.5ex
\def\baselinestretch{1.50}
\parskip 5pt plus 1pt

\begin{abstract}
Unification ideas motivate the formulation   of field equations on
an extended spin space.
 Demanding that the Poincar\'e symmetry be
maintained, one derives scalar symmetries that are associated with
flavor and gauge groups.  Boson and fermion solutions are obtained
with a fixed representation. A  field theory can be equivalently
written and interpreted in terms of elements of
 such
space and is similarly constrained. At 5+1 dimensions, one obtains
isospin and hypercharge  $SU(2)_L\times U(1)$ symmetries, their
vector carriers,
two-flavor charged and chargeless leptons, and scalar particles. 
Mass terms produce  breaking of the symmetry to an electromagnetic
$U(1)$, a Weinberg's angle with $sin^2(\theta_W)=.25$, and
additional information on the respective coupling constants. Their
underlying spin symmetry gives information on the particles'
masses;  one reproduces the standard-model ratio $M_Z/M_W$, and
predicts a Higgs mass of $M_H\approx 114$ GeV, at tree level.

\end{abstract}
\vskip .5cm

\centerline{PACS: 12.10.-g, 04.50.+h, 14.80.Cp, 03.70.+k }


\vskip 1cm


\baselineskip 22pt \vfil\eject \noindent
 \section {Introduction}

Although the accepted theory of elementary particles, the standard
model (SM),
 is
successful in describing their behavior, it is phenomenological;
  a justification   for its input  is  still required at a more fundamental level,
as evidenced by the large number of experimental parameters it
needs. We also lack information on the origin of the particular
groups $SU(3)\times SU(2)_L\times U(1)$ which define the
interactions, and on why the isospin acts only  on a given
chirality. Neither is it clear the origin of the spectrum of
fermions, the three generations, the masses and mixing angles, and
the gauge-group fundamental representations in which they appear.
The Higgs particle is a useful mathematical device but we lack a
more basic reason for its presence.

The idea of unification has proved to be a powerful tool in
physical research, finding links between originally assumed
separated phenomena, and also predicting new ones. For example, by
assuming the SM interactions have a shared origin in the same
unified group\cite{unification} some clues can be obtained on the
fermion representations in the SM and also relations among the
coupling constants.


On another plane, the Kaluza-Klein idea, originally used in the
framework of general relativity, has been applied to understand
interactions by relating them to additional spatial dimensions.
Other unification ideas may shed  more light on particle-physics
problems. Spin is a physical manifestation of the
  fundamental representation of the Lorentz group
 and it is more so   in relation to
 space,  which uses the vector representation.
While both the spin and configuration spaces are equally necessary
elements for the description of particles, the  association of
interactions to extended spin spaces is feasible, and gives rise
to a model with new connections to the
SM\cite{Jaime},\cite{Jaimenext}.

  In this paper we give a field-theory formulation of the model, providing
  it with additional fundament, and we further explore  the 5+1 dimensional case.
  We  review the formulation of
Poincar\'e-invariant field equations on an extended 5+1 d spin
space, and the link of its symmetries and solutions to the SM
electroweak  sector. We also show that a  field theory  can be
equivalently formulated in such a space and is similarly
constrained. Thus, the demand that fields be in a 5+1 dimensional
spin space determines the interactions and  representations of the
SM electroweak  sector, and additional constraints. In particular,
we extract information on the particle masses, using the
underlying spin symmetry, and making a comparison with the SM.

We first consider an extended Dirac equation and develop its
surrounding formalism, using the conventional relativistic
quantum-mechanical framework (Section 2). We show that  boson
(Section 3) and fermion (Section 4) solutions are obtained that
share
 the same spin solution space.     The simplest extension of the equation at 5+1 dimensions
predicts an $SU(2)_L\times U(1)$ symmetry.  We obtain solutions
with quantum numbers of the corresponding gauge fields, leptons,
and scalars interacting electroweakly (Section 5).  Typical
Lagrangians are
 equivalently written in terms of the above degrees of freedom,
with the dimension
  determining also
the representations and the interactions (Section 6).
 A mass term set with an underlying  spin symmetry predicts boson
 masses  as in the  SM and  the other particles' (Section 7).
We also obtain information on the electroweak SM  coupling
constants (Section 8).


\section{ Generalized Dirac equation}

We depart from the equation
\begin{eqnarray}
\label {Jaimeq} \gamma_0( i \partial_\mu\gamma^\mu -M)\Psi ={ 0},
\end{eqnarray}
where, with the aim of jointly describing  bosons and fermions, we
now assume $\Psi$ represents a matrix. Eq. \ref{Jaimeq} contains
four conditions over four spinors in a $4\times 4$ matrix.
 There are, then, additional possible ones to further classify $\Psi$.
 The  transformations and symmetry operations on the Dirac
operator  $( i \partial_\mu\gamma^\mu -M)\rightarrow {U}(\it  i
\partial_\mu\gamma^\mu -M) {U^{-1}}$
 induce
the lhs of the transformation
\begin{eqnarray}
\label  {transfo}
 \Psi\rightarrow U \Psi U^\dagger.
\end{eqnarray}
 Also, $\Psi$ is  postulated to transform  as indicated on the rhs.
  The latter transformation  is consistent with the additional equation
\begin{eqnarray}
\label {Jaimeqnext} \Psi\gamma_0( -i \stackrel{\leftarrow}{
\partial_\mu}\gamma^\mu -M) ={ 0}
\end{eqnarray}
(the Dirac operator  transforming accordingly).\footnote{The
Bargmann-Wigner equations\cite{Bargmann}  contain Eq. \ref{Jaimeq}
but set the different second condition as $\Psi^*_{BW}\gamma_0 (-i
\stackrel{\leftarrow}{\partial_\mu}\gamma^\mu-M)={ 0}$
$\Psi^*_{BW}$ contains charge conjugated $\langle w_j |$
components (see below). } In fact, the conjugated fields
$\Psi^\dagger$, for which Eq. \ref{transfo} is valid, satisfy this
equation. It is by taking also account of this kind of fields that
we can span the function space within the 32-dimensional complex
$4 \times 4$ matrices. Then, we shall extend our space of
solutions by considering also combinations of  fields ${ A}+{
B}^\dagger$.

$\Psi$ can be understood to be constructed of  a tensor product of
the usual configuration (or momentum) space, and a column $|w_i
\rangle$ and a row $\langle w_j |$ state that can have a spinor
interpretation\cite{Lifshitz},  with expansion $\sum_{i,j} a_{ij}
|w_i \rangle\langle w_j | $. The resulting tensor product of
operators  acting on this tensor-product space and classifying the
solutions  consist generators of the Poincar\'e algebra with spin
components, (or scalars) acting as $U$
 on each side of $\Psi$, and  derivative ones
acting only once, from either side;  products of derivative and
spin operators act from both sides.
   The expansion of
$\Psi$  implies it Lorentz transforms  as a scalar, a vector, or
an antisymmetric tensor. In fact, we will show below certain
choices of symmetry operators allow for a fermion interpretation
of some solutions too.

 If ${ A}$, ${ B}$ are
solutions, the matrix product ${ C}={ AB}$, defines an algebra
whose elements may or may not be solutions. The inner product of
${ A}$, ${ B}$  is naturally defined  by
\begin{eqnarray}
\label{pointproduct}
   \langle  { A} | { B} \rangle =tr({ A^\dagger B}),
\end{eqnarray}
as expected for  this extended  tensor-product space. A trace over
the coordinates is also implied.

We are interested in plane-wave solutions of the form
\begin{eqnarray}
\label {planewaves}
{ \Psi}^{(+)}_{ki}(x)&=&{ u_i}(k)e^{-i kx}\\
{ \Psi}^{(-)}_{ki}(x)&=& { v_i}(k)e^{i kx}, \label {planewavesend}
 \end{eqnarray}
where $k^\mu$ is the momentum  four-vector $(E,\bf k)$, $k_0=E$,
and $u_i$, $v_i$ are matrices containing information on the
polarization.
As applied to  ${ \Psi}^{(+)}_{ki}(x)$ in Eq. \ref{planewaves},
 Eq. \ref{Jaimeq} defines the Hamiltonian  $H_{lhs}=\gamma_0(  {\bf k }\cdot{\boldmathgamma}+ M)$  and a projection of the
Pauli-Lubansky vector on the  four-vector $n_k=(1/M)( | {\bf k}|
,E \hat{{\bf k}}   )$, $(1/M)W\cdot n_k={\bf \Sigma}\cdot{\bf \hat
k  }$, which uses
\begin{eqnarray}
\label {angular}
 J_{\mu\nu}=i(x_\mu\partial_\nu-x_\nu\partial_\mu)+\frac{1}{2}\sigma_{\mu\nu},
\end{eqnarray}
with $\sigma_{\mu\nu}=\frac{i}{2}[ \gamma_\mu,\gamma_\nu ]$   and
the spin operator ${\bf
\Sigma}=\frac{1}{2}\gamma_5\gamma_0{\boldmathgamma}$, which
 is valid both for the massless and the massive cases.
However, ${ \Psi}^{(+)}_{ki}(x)$  is classified by the
energy-momentum symmetry operators corresponding to Eq.
\ref{Jaimeqnext} (always in the relativistic quantum-mechanics
framework) as a negative-energy solution. A consistent
characterization  of the latter can be given if we assume
 its rhs $\langle w_j |$ spinor component  to be a hole.
This  interpretation is set  for operators acting  on $\Psi$ from
the rhs, as well as  on the $\Psi^\dagger$ fields. Thus, the
hermitian conjugate of   ${{ \Psi}^{(-)}_k}^\dagger(x)$  in Eq.
\ref{planewavesend} has the same exponential dependence as ${
\Psi}^{(+)}_k(x)$ in Eq. \ref{planewaves}, and is also interpreted
as a positive-energy solution.

\section{ Boson solutions}
On Table  \ref{VmA} are shown solutions of the massless equation
\ref{Jaimeq}, bilinear in the $\gamma$s, and  they are given
together with their quantum numbers. We give them in terms of
their polarizations, setting the coordinate dependence as in Eq.
\ref{planewaves}, and we assume the spatial component of $k^\mu$
to be along $\hat {\bf z}$.
Here and throughout the solutions are normalized as $tr( \bfu
^\dagger_ {i}( {k)} \bfu _{i}( {k)}) =1$. In the first two
columns, we  present the eigenvalues $\lambda$ of the operators
$O$ in the form $O{ u}_i({k} )= \lambda { u}_i({k} )$.

The solutions on Table \ref{VmA} are interpreted as   vector
bosons, for
 Eq. \ref{Jaimeq}   alone  implies they satisfy the
Klein-Gordon equation and  ${\bf \Sigma\cdot{\hat{\bf  k} }}$
classifies their  different degenerate polarization components
 as vectors.  According to the direct-product operator convention, the helicity operator  is calculated
in the fourth column with ${\bf\Sigma\cdot{\hat{\bf  k}
}}_{lhs}\Psi+\Psi {\bf \Sigma\cdot{\hat{\bf  k} }}_{rhs}$, where
${\bf\Sigma\cdot{\hat{\bf k} }}_{rhs}$, and its (rhs) eigenvalue
are unchanged under the hole interpretation. The use of a
commutator follows from the equality ${\bf\Sigma\cdot{\hat{\bf  k}
}}_{lhs}\Psi+\Psi {\bf\Sigma\cdot{\hat{\bf  k}
}}_{rhs}={\bf\Sigma\cdot{\hat{\bf  k} }}_{lhs}\Psi-\Psi
{\bf\Sigma\cdot{\hat{\bf  k}
}}_{lhs}\equiv[{\bf\Sigma\cdot{\hat{\bf  k} }},\Psi]$. The third
column characterizes the solutions with the application of
$H_{lhs}\Psi+\Psi H_{rhs}$, with $H_{rhs}$ coming  from Eq.
\ref{Jaimeqnext}.   Although the rhs spinor of $\bfu_{-1}({k})$ is
not on-shell, as measured by $H_{rhs}$,  the latter gives the
expected energy sign in the hole interpretation.
Within   these  rules, $\bfu_{-1}({k})$, $\bfu_{-1}({\tilde k})$
are  on-shell particles with transverse polarizations, with
helicity $-1$, propagating respectively in the $\hat {\bf z}$ and
$-\hat {\bf z}$ directions;  the latter is denoted through the
four-vector $\tilde  k^\mu=k_\mu$;  $\bfu_{0}({k})$, $\bfu_{0}(
{\tilde k} )$ are off-shell and polarized in the
longitudinal-scalar directions. The $u_i$ solutions  do not
represent independent polarization components as, e.g.,
$\bfu_{i}({\tilde k})$ can be obtained by rotating the
$\bfu_{i}({k})$ (applying a Lorentz transformation of the form
\ref {transfo}).


In the massless case, we have also negative-energy solutions ${
v}_{ i}({k} )=\bfu_{i}(k)$ (and $\tilde k$ terms) from  Eq.
\ref{planewavesend},
 that is, with opposite helicities.
 The positive-energy solutions containing ${v}^\dagger (k),$ ${ v}^\dagger (
\tilde k)  $ have also  opposite helicities. By applying the
parity operator $\gamma_0 \wp  $, with $ \wp x_\mu=\tilde x_\mu $,
in the form \ref{transfo} one classifies the solutions according
to the weight of the  vector $V$ and axial $A$ components. The
solutions in Table \ref{VmA} are $V-A$, while $V+A$ solutions are
also obtained. Their combination,
\begin{eqnarray} \label{Amufirst} { A}_\mu(x)
=g_{\mu\nu}(x)\frac{i}{2}\gamma_0\gamma^\nu ,
\end{eqnarray}
 with $g_{\mu\nu}(x)$ a  polarization tensor,
 transforms  into $ { A}^\mu ({\tilde x})$ under parity $P$, that is, as a vector.
An axial term can be also obtained. A combination of  ${
A}_\mu(x)$ and its hermitian conjugate can be shown to transform
as a vector under the charge conjugation operator $C=i\gamma_2
\mathcal{K} $, with  ${\mathcal{K}} i = - i  {\mathcal{K}} $;
given its quantum numbers,
 it becomes then possible to relate ${ A}_\mu(x)$  to
the vector potential of an electromagnetic field satisfying
Maxwell's equations within the Lorentz gauge. Actually, we may
also view $\frac{1}{2}\gamma_0\gamma_\mu$ as  an orthonormal
polarization
 basis, $A_\mu=tr\frac{1}{2}\gamma_\mu A^\nu\frac{1}{2}\gamma_\nu$\footnote{As for
 $\bar \psi=\psi^\dagger\gamma_0$, a unitary transformation can be
 applied to
 the fields and operators to convert them to a covariant form.}; just as
   $n_\mu$ in
  $A_\mu=g_{\mu\nu}A^\nu =n_\mu\cdot A^\nu n_\nu$.
In fact, the sum of  Eqs.  of \ref{Jaimeq} and
 \ref{Jaimeqnext} implies   for a $\Psi$ containing $\gamma_0\slash\!\!\!\! A={ A}^\mu\gamma_0\gamma_\mu $  that  ${
 A}^\mu$ satisfies the free Maxwell's equations$\cite{Bargmann}$. Both interpretations
for  the polarization components as basis for solutions in an
extended Dirac equation, and for standard fields shall be
considered in this work.

The remaining eight degrees of freedom   in the massless case
contain
 six forming an
antisymmetric tensor ${ A } _{\mu\nu}=\frac{1}{4} \gamma_0[\gamma_
\mu,\gamma_ \nu], $   and scalar, and pseudoscalar terms ${\phi
}=\frac{1}{2} \gamma_0$, ${\phi _5}=\frac{1}{2} \gamma_5\gamma_0$.

\section{ Fermion  solutions} The chirality invariance allows for more choices of Lorentz generators.
Using $J_{\mu\nu}^-=\frac{1}{2} (1-\gamma_5) J_{\mu\nu}$ in  ${\bf
\Sigma}$ to classify the  solutions in  Table \ref{VmA},
 they remain $V-A$.
\begin{table}[h]
\begin{eqnarray}
\label{lefthand} \nonumber Vector\ solutions & \gamma_0\gamma^3 &
\frac{i}{2} \gamma_1\gamma_2 \ \ [ H/k_0,\ ] \ \ [{\bf
\Sigma\cdot{\hat{\bf  p} }} ,\ ] \\ \nonumber \bfu_{-1}({k})
=\frac{1}{4} (1-\gamma_5)\gamma_0(\gamma_1-i \gamma_2)& 1 & -1/2\
\ \ 2\ \ -1\\ \nonumber \bfu_{-1}({ \tilde{k}}) =\frac{1}{4}
(1-\gamma_5)\gamma_0(\gamma_1+i \gamma_2)&  -1 & 1/2\ \ 2 \ \ -1\\
\nonumber \bfu_0( {k})= \frac{1}{4}
(1-\gamma_5)\gamma_0(\gamma_0-\gamma_3)& 1 & -1/2\  \ \ \ 0 \ \ \
0 \\ \nonumber \bfu_{0}( {\tilde  k)} =\frac{1}{4}
(1-\gamma_5)\gamma_0(\gamma_0+\gamma_3)& -1 & 1/2 \ \ \ 0\ \ \ 0
\nonumber
\end{eqnarray}
\caption{\label{VmA}V-A\   terms. }
\end{table}
 However, when applying $J_{\mu\nu}^-$ to the other solutions of the form
$(1\pm\gamma_5)\gamma$   it leads to one of the sides (either $|
w_i \rangle $ or $\langle w_j | $) to transform trivially,   and
therefore, to spin-$1/2$ objects transforming
 as the $(1/2,0)$ or $(0,1/2)$
 representations  of the  Lorentz group.
Having obtained   both fermion  and boson  solutions constitutes
progress in the  task of giving a unified description of these
fields. Clearly, their nature depends on the Hamiltonian and on
the set of valid symmetry  transformations, chosen among limited
options. But once the  choice is made, there is no ambiguity.

The equation    \begin{eqnarray} \label {Jaimeferq} i(1-\gamma_5)
\gamma_0\partial_\mu\gamma^\mu  {  \Psi}=0
\end{eqnarray}
has all these fields  as solutions. If we stick to $J_{\mu\nu}^-$
to classify them, the invariance algebra of the equation
 has  as  additional symmetry the group of linear complex transformations $G(2,C)$
with eight components, generated by $(1/2) (1+ \gamma_5)$ and
$f_{\mu\nu}=-(i/2)   (1+\gamma_5) \sigma_{\mu\nu}$. The unitary
subgroups $SU(2)\times U(1)$  of   $G(2,C)$ imply two additional
quantum numbers we can assign to the  solutions. In consideration
that  this symmetry does not act on  the vector solution part, and
taking account of the known quantum numbers  of fermions  in
nature,  we shall  associate these
 operators with the flavor and spin-1/2 number operators,
respectively. The  $SU(2)$ set of operators leads   to a flavor
doublet.  The $U(1)$ is in this case not independent from the
chirality. Choosing $f_{30}$ to classify the solutions of Eq.
\ref{Jaimeferq}, these  are given in Table \ref{fermionsend} (as
$H$ and ${\bf \Sigma}\cdot{\bf \hat p  }$ act trivially from the
rhs, commutators are not needed.) Overall, the matrix objects we
use allow for a fermion interpretation, for one of its components
behaves as a spinor, while the other carries  the flavor quantum
number.

\begin{table}[h]
\begin{eqnarray}
Left-handed\  spin\ 1/2 \  particles &  \frac{1}{2}(1-\gamma_5)\gamma_0\gamma^3 & \frac{i}{4} (1-\gamma_5)\gamma_1\gamma_2  \ \ [f_{30},\ ] \nonumber \\
\bfw_{-1/2}({ k})  = \frac{1}{4} (1-\gamma_5) (\gamma_0+\gamma_3 ) & 1 & -1/2\ \ 1/2 \nonumber\\
\bfw_{-1/2}  ({\tilde   k})= \frac{1}{4} (1-\gamma_5)  (\gamma_1+i\gamma_2 )& -1 & 1/2\ \ 1/2 \nonumber \\
\hat \bfw_{-1/2}   ({ k})= \frac{1}{4} (1-\gamma_5 )
(\gamma_1-i\gamma_2 )& 1 & -1/2\  -1/2\nonumber\\ \nonumber \hat
\bfw_{-1/2}({ \tilde   k})  =  \frac{1}{4} (1-\gamma_5)
(\gamma_0-\gamma_3 ) & -1 & 1/2\ \ -1/2
\end{eqnarray} 
\caption{Massless  fermions. \label{fermionsend}}
\end{table}

\section{ 5+1 Dimensional Extension}
The simplest generalization of the above model is to consider the
six-dimensional Clifford algebra, (the $d=5$ lives also in a
$4\times 4$  space),
 composed of  64 $8\times 8$  matrices. To describe it we use
 the quaternion-like objects $1$, $I$, $J$, $K$ and now generalized  $8 \times 8$ matrices $\gamma_\mu$ (and $\gamma_5$).
A  6-$d$ Clifford algebra $\{ \gamma_\mu^\prime,\gamma_\nu^\prime
\}=g_{\mu\nu}$ can be formed with   the   $4$-$d$ elements
 $\gamma_\mu^\prime=\gamma_\mu,\ \ \mu=0,1,3,$ $\gamma_2^\prime=I \gamma_2$, and the 4-$d$ scalars
$\gamma_5^\prime=  J \gamma_2,\  \gamma_6^\prime=  K \gamma_2$.
Altogether, the scalars (and pseudo-) are $1$, $I$, $J \gamma_2$,
$K \gamma_2$, $\gamma_5$, $I\gamma_5$, $J \gamma_2\gamma_5$, $K
\gamma_2\gamma_5$. From these,  $I\gamma_5$ commutes with the
rest. Excluding it and  the identity, the remaining six elements
generate an $SO(4)$ algebra, or equivalently, an $SU(2)\times
SU(2)$ algebra. The eight scalars have a Cartan algebra of
dimension four, for which we can take the basis $1$, $I$,
$\gamma_5$, $I\gamma_5$.

 We obtain the useful prediction of a limited number of
scalar symmetries and representations consistent with the 4-$d$
Lorentz symmetry. We may choose among them the closest to
reproduce aspects of the SM. The only projector operator $L$ that
can be constructed with the latter operators, describes both
fermions and bosons through the Lorentz generator
$J_{\mu\nu}^L=LJ_{\mu\nu}^\prime$, with $J_{\mu\nu}^\prime=
i(x_\mu\partial_\nu-x_\nu\partial_\mu)+\frac{1}{2}\sigma_{\mu\nu}^\prime$,
$\sigma_{\mu\nu}^\prime=\frac{i}{2}[\gamma_\mu^\prime,\gamma_\nu^\prime]$,
allows for a non-abelian group symmetry, and permits parity to be
a good quantum number for some solutions is\cite{Jaime}
$L=\frac{3}{4}-\frac{1}{4}(I+\gamma_5+I \gamma_5
)=1-\frac{1}{4}(1+I)(1+I \gamma_5)$; the latter can be interpreted
as the lepton number. The equation
\begin{eqnarray}
\label {geneqIII} i L\gamma_0  \partial^\mu\gamma_\mu^\prime
\Psi=0,\ \mu=0,...,3,
\end{eqnarray}
is invariant under $J_{\mu\nu}^L$, and allows for a flavor
symmetry as defined previously. It is also invariant under
 the $L$-commuting  scalars: the
 hypercharge $Y=-1+1/2(I+\gamma_5),$ and the isospin $SU(2)_L$ with generators
 \begin{eqnarray}
 \label {iso}
 I_1&=&\frac{i}{4}(1-I\gamma_5)J\gamma_2 ,\\
I_2&=&-\frac{i}{4}(1-I\gamma_5)K\gamma_2,\\
I_3&=&-\frac{1}{4}(1-I\gamma_5)I \label {isolast};
\end{eqnarray} this  identification of $Y$ and
$I_i$ comes from their commuation relations, and  from the correct
vertices and quantum numbers they determine on the bosons and
leptons, as argued below. $Y$ can also be deduced from other
requirements\cite{Jaime} related to gauge symmetry.

These  scalars  lead, as expected,   to a restricted set of
solutions. The massless positive  solutions are shown
schematically     on Table \ref{tab:tablejb} (see  Ref.
\cite{Jaime} for details) and they include $V-A$ and $V$+$A$
vectors, with  components
$B_\mu=\frac{1}{4\sqrt{2}}(1-I)(1+I\gamma_5)\gamma_0\gamma_\mu,$
$\tilde B_\mu=\frac{1}{4}(1-I\gamma_5)\gamma_0\gamma_\mu$, which
amount to eight degrees of freedom (df); isospin  $V-A$ vectors
$W_\mu^i=I_i \tilde B_\mu$ with twelve df; also left and
right-handed antisymmetric tensors and scalars $n_{L,R}$,
$v_{L,R}$ in an isospin doublet, amounting to eight df, and with
their antiparticles sixteen df. These add up to thirty-six bosons.
We  also obtain spin-1/2 left-handed particles in an isospin
doublet $(\nu,l^-)_L$ with $Y=-1$ and a right-handed isospin
singlet  $l_R^-$ with $Y=-2$; for example
\begin{eqnarray}
\label{lR} l^-_{R}(k)= \frac{1}{8} (1+I \gamma _5 ) ( J \gamma_2-i
K\gamma_2) \gamma_0 (\gamma_1+ i I \gamma_2) \end {eqnarray}
 is a right-handed spin component with
these quantum numbers and given flavor. Taking account of
antiparticles and the two flavors, we have twenty-four fermion df.
 The reason for not having altogether sixty-four active df is the four
inert df, defining   the flavor, projected by $1-L$,  and  which
are not connected to  the Hamiltonian (top-left matrix on Table
\ref{tab:tablejb}).

\renewcommand{\arraystretch}{3.5}
\begin{table}[h]
\center{ \begin{tabular}{|c|c|c|c|} \hline
   & $ {\Large e^+_L} $ & $\  \ \ \bar\nu_ {R} \ \ \ $  & $l^+_R$    \\ \hline
\ \ \ \ $e^-_R$\ \ \ \ & $\ \ \ \tilde B_{\mu} \ \ \   $  &
$\tilde n_R$ & $\tilde v_R$  \\ \hline $ \nu_L$   & $ \ \ \ n_L$\
\ \ & $B_{ \mu},\ W^0_\mu$ & $W^1_\mu ,\   W^2_\mu$  \\ \hline
 $l^-_L$     & $v_L$ & $W^1_\mu ,\   W^2_\mu$   & $B_{\mu},\   W^0_\mu$ \\ \hline
\end{tabular}}
\caption{\label{tab:tablejb} Arrangement of solutions in a $6-d$
$8\times 8$ matrix model, with each box occupying a $2\times 2 $
matrix.}
\end{table}
\renewcommand{\arraystretch}{1}
\normalsize

In seeking a massive extension of Eq. \ref{geneqIII}, we expect
all the hermitian combinations of the
 scalar terms $M_i\gamma_0$
 to be scalars with respect to
  $J_{\mu\nu}^\prime$.
However, if we also demand that they be scalars  with respect to $
J_{\mu\nu}^L,$ then the choices are reduced to $M_1 = (M/2)
(1-I)$,   $M_2 =  i(M/2) (\gamma_5-I \gamma_5)$,
 $M_3=-( M /2)J\gamma_2  (1+\gamma_5)$,   $M_4 =  (M/2 )K\gamma_2  (1+\gamma_5)$,
  where $M$ is the mass constant. Now, the only non-trivial scalar that commutes with   all  $M_i\gamma_0$ terms is $L$.
Nevertheless, if we relax this condition we obtain in addition
that  only
\begin{eqnarray}
\label{charge} Q=I_3+\frac{1}{2} Y
\end{eqnarray}
commutes with $M_3\gamma_0$ and $M_4\gamma_0$. As $Q$ is the
electric charge we deduce the electromagnetic $U(1)_{em}$ remains
a symmetry while the hypercharge and isospin  are broken. We
stress that   $Q$ is deduced, rather than being imposed, as the
only  additional symmetry consistent with massive terms.

\section{ Interactive field theory and physical fields}
We argue below that an interactive field theory can be constructed
for fields belonging to an extended spin space. Although we refer
to 5+1 $d$, this applies to other dimensions as well. As in
Section 5, the spin-space dimension determines the
representations, and we shall also  find constraints on their
possible interactions.

 The expression for the  kinetic-component  Lagrangian density
  of a non-abelian gauge-invariant vector field
\begin{eqnarray}
\label{kineLagra}
  { \mathcal
L}_{V}= -\frac{1}{4}F_{\mu\lambda}^ag^{\lambda\eta}\delta_{ab}
F^{b\mu}_{\ \  \eta}=-\frac{1}{4 N_o}tr { \mathcal P}_{D}
F_{\mu\lambda}^a\gamma_0\gamma^{\lambda} G_a F^{b\mu}_{\ \
\eta}\gamma_0\gamma_ \eta G_b
\end{eqnarray}
 shows ${\mathcal   L}_{V}$
is equivalent to a trace over combinations over normalized
components $\frac{1}{\sqrt{ N_o}}\gamma_0\gamma_\mu G_a, $
$\mu=0,...3,$ with coefficients $F^a_{\mu\nu}=\partial_\mu
A_\nu^a-
\partial_\nu^a A_\mu^a+ g A_\mu^b A_\nu^c C^a_{bc}$, $g$ the
coupling constant,   $C^a_{bc}$ the structure constants, and ${
\mathcal P}_{D}$ a Lorentz-scalar projection operator, defining
the solution space and Poincar\'e algebra in 5+1 $d$. Limited
choices exist for ${ \mathcal P}_{D}$, and the restrictions in
Section 5 imply ${ \mathcal P}_{D}=L$, so,
 {\it e. g.}, $G_a=I_a $  are $8\times 8$ matrices,
  and $N_o g^\mu_{\ \nu}\delta_{ab}=tr\gamma_0\gamma_\mu G_a \gamma_0\gamma_\nu G_b=g^\mu_{\ \nu}trG_a G_b $, where for
non-abelian irreducible representations we use $trG_i G_j=2
\delta_{ij}$.

  Similarly,   the interactive part of the fermion gauge-invariant Lagrangian
\begin{eqnarray}
\label{inteLagra}
  {\mathcal   L}_{f}=\frac{1}{2}{\psi^\alpha}^\dagger\gamma_0(
i\stackrel{\leftrightarrow}\partial_\mu-g A^a_\mu G_a
)\gamma^\mu\psi^\alpha,
\end{eqnarray}
 with $\psi^\alpha$  a  massless spinor with flavor $\alpha$,
can be written
\begin{eqnarray}
\label{inteLagratra}
    {\mathcal   L}_{int} = -g\frac{1}{ 2 N_o} tr
L A_{\mu}^a \gamma_0\gamma^\mu G_a j^{b\alpha}_\lambda
\gamma_0\gamma_\lambda G_b,\end{eqnarray} with  $
j_\mu^{a\alpha}=tr{\Psi^\alpha}^\dagger\gamma_0\gamma_\mu G_a
\Psi^\alpha$
 containing $\Psi^\alpha=\psi^\alpha\langle \alpha |$, and $\langle \alpha|$
 is a row  state accounting for the flavor. Similar matches between standard Lagrangians
 and  their expressions in terms of fields in an extended spin space
 for components with scalar fields, and their
 interaction with the vector and spin-1/2 fields,  are
 considered below.
 ${\mathcal   L}_{int}$ presents $A^a_\mu$ and $j^a_\mu$
 as components over $\frac{1}{\sqrt{ N_o}}\gamma_0\gamma_\mu G_a, $
 that is, the vector field and the current occupy   the same spin space. This
connection and the quantum field theory (QFT) understanding of
this vertex as the transition operator between fermion states,
exerted by a vector particle, with the coupling constant as a
measure of the
 transition probability,  justifies the  interpretation  for it $\frac{1}{2}g A^{a\mu} j^{a\alpha}_\mu= A^{a\mu}\frac{1}{\sqrt{ N_o}}
tr{\Psi^\alpha}^\dagger \gamma_0\gamma_\mu G_a\Psi^\alpha$,
leading to the  identification $g\rightarrow 2\sqrt{ \frac{K}{
N_o}}$, $K$ correcting for  over-counted  reducible
representations, which need not be considered at 5+1 $d$. The
theoretical assignment of $g$ complements QFT, in which the
coupling constant
 is set
 experimentally.
  It should be also
understood as tree-level information, while the values  are
modified by the presence of a virtual cloud of fields, at given
energy. Although in QFT the coupling constant
 is obtained  perturbatively in terms of powers of
 the bare, which takes  infinite values  absorbed through renormalization,
 we may take the view that renormalization is a
calculational device and that its physical value  is a
manifestation of the bare one;
 this is feasible for  small  coupling constants,  which  can give
 small corrections. Energy corrections are also necessary for a more detailed
calculation.

  We  found
in Section 5 that among a limited number of choices, at 5+1, $L$
reproduces  SM input. In this section's approach, it restricts the
possible gauge symmetries and representations,  for $
\gamma_0\gamma_\mu G_i$ needs to be contained in
 the space it projects. Thus,
it determines the symmetries,
   which are global, and  in turn,
 the  allowed gauge interactions. Furthermore, it
 fixes the representations. In the same vein,
 it
restricts the allowed Lagrangians.
 Fixing $L$, the   physically feasible
($trG_i G_j=2 \delta_{ij}$) gauge groups are
comprised$\cite{Weinbergbook}$ of the $U(1)$ and compact simple
algebras generated in it.  With the conditions of
 renormalizability (this excludes antisymmetric fields),  and $L$ conservation, the
allowed Lagrangian reduces basically to that of the electroweak
sector of the SM, with representations as on Table
\ref{tab:tablejb}.



\section{ Boson fields after spontaneous symmetry breaking and their masses }

Information on SM  fields and their masses is next derived from
the polarization and generator
 components of the 5+1 $d$ spin solutions and their
 expectation values under  mass operators and then, in relation to the SM.
 These values' scalar nature must be
 ensured in order to describe mass.

In  the SM,
 quadratic
terms of the form $\langle  { \bar \phi} |F^\dagger F |{ \bar\phi}
\rangle$, where $\bar\phi$ is a scalar-particle field (and
eventually, a vacuum expectation value), give masses to the boson
fields $F$. A similar scalar-expectation value  for bosons is
implemented using Eq. \ref{pointproduct}  by
\begin{eqnarray}
\label{ratioM} M_F^2=tr [ \tilde H,F]^\dagger_\pm [\tilde
H,F]_\pm,
\end{eqnarray}
where $\tilde H$ is constructed with  the  scalar-chargeless
solutions of $n_{L,R}$, which are the counterpart of the Higgs
particle in the SM. Indeed, these terms' components $\tilde
n_0=\frac{1}{4 \sqrt{2}}(1+I\gamma_5)(J\gamma_2-i K
\gamma_2)\gamma_0=\frac{i}{ \sqrt{2}}\gamma_0 I_+$, $
n_0=\frac{i}{ \sqrt{2}}\gamma_0I_-$, $I_\pm=I_1\pm iI_2$, from
Eqs. \ref{iso}-\ref{isolast},
  form the parity-conserving scalar
combinations $M_i\gamma_0$, $i=3,4.$ Specifically,
\begin{eqnarray}
\label{ratioMvscale} \tilde  H= v^\prime \frac{1}{\sqrt{2}}(n_0+i
\bar n_0)
\end{eqnarray}  with $v^\prime$  an energy scale, is symmetric in particle and antiparticle $n_0$,
$\bar n_0$ components and we justify  the coefficients' choice
below. Covariance requires  the use of anticommutators   for the
spatial vector components, leading to the same scalar quantity,
hence the $\pm$ index in Eq. \ref{ratioM}. When applying this
equation to the  scalar fields  both methods give the same result.

We note   that  $M_i\gamma_0,$ $i=3,4,$ by construction,
immediately give  $[n_0 , Q ] =[\bar n_0 , Q ]=0$. Thus, the term
$A_\mu=\frac{1}{2}Q\gamma_0\gamma_\mu$ has quantum numbers of a
massless vector and induces the  fermion vertex that identifies it
with a photon.


 There are two additional neutral combinations of vector bosons
orthogonal to $Q$. One, the axial combination $Z_\mu$ can be
related to its namesake in the SM. Indeed, $A_\mu$ can be
represented as a mixture of two chargeless and massless
components.
 On the  one hand,  the $B_\mu$  and  $\tilde B_{\mu}$ terms form
$B_{Y\mu}   =\frac{1}{2\sqrt{3}}Y \gamma_0\gamma_\mu$ ($B_{Y0}
=g^\prime \frac{1}{2}Y$),
 that is, the hypercharge carriers.
 Thus,
we obtain another  argument   to set $Y$ whose origin is  in the
way we arrive at the expression for $Q$ in Eq. \ref{charge}. With
hindsight, we write in parenthesis  the $B_{Y0}$ component, and
other fields' below, more generally in terms of a hypercharge
coupling constant $g^\prime$ which encompasses  both the arbitrary
parameter and normalized-linked interpretations.
On the other hand, we can use the chargeless vector isospin
triplet component
$W_\mu^3=  I_3\gamma_0\gamma_\mu$ ($W_0^3=  gI_3$), $g$ the $SU(2)$ coupling. 
From the expressions for $Q$, $A_\mu$, and  $W_\mu^3$  we easily
obtain
\begin{eqnarray}
\label{Weinangle}
A_\mu=\frac{1}{2}W_\mu^3+\frac{\sqrt{3}}{2}B_{Y\mu}.
\end{eqnarray}
The value of Weinberg's angle $\theta_W$  is derived  immediately
from this equation by making an analogy with the  combination of
fields obtained in the  SM, after application of the Higgs
mechanism\cite{Weinberg}. The photon there has the form
\begin{eqnarray}
\label{photon}
  \bar A_\mu=\frac{1}{\sqrt{g^2+g^{\prime 2}}}(g\bar B_{Y\mu}+g^\prime \bar
  W^0_\mu),
  \end{eqnarray}
where the bar denotes here and below SM fields. We obtain
$\frac{g^\prime}{g} =\frac{1}{\sqrt{3}}$. As in the SM
$tan(\theta_W)=\frac{g^\prime}{g}$, we find $sin^2(\theta_W)=.25$
. We derive the other possible combination constructed with this
form:
\begin{eqnarray}
\label{Zform} Z_\mu=\frac{\sqrt{3}}{2}W_\mu^3-\frac{1}{2}B_{Y\mu}
\end{eqnarray}
($Z_0=\frac{1}{\sqrt{{g}^2 +{g^\prime}^2}}({g^\prime}^2
\frac{1}{2}Y-g^2 I_3)$).

  To calculate $M_F$ in Eq. \ref{ratioM}  for the vector bosons   we may use  the underlying isospin
and hypercharge $SU(2)_L\times U(1)_Y$ classification of the
generators. It gives $[\tilde n_0,Z_0]=\frac{1}{2}\sqrt{{g}^2
+{g^\prime}^2}\tilde n_0$, $[ n_0,Z_0]=-\frac{1}{2}\sqrt{{g}^2
+{g^\prime}^2}  n_0$. For the charged vectors,
$[n_0,W_0^+]=\frac{g}{\sqrt{2}} v^+_0$, $[\tilde n_0,W_0^+]=0,$
where $\tilde v_0^-=\frac{1}{4
\sqrt{2}}(1+I\gamma_5)(1+I)\gamma_0$, $  v_0^+=\frac{1}{4
\sqrt{2}}(1-I\gamma_5)(1+I)\gamma_0$  are  charged-scalar
components.
 The resulting values are presented on Table
\ref{tab:tablejbn}, in terms of the $W^\pm_\mu$ mass. They are
compared with SM expressions and values. Remarkably, the
calculation reproduces the SM
 relation $M_Z/M_W=\sqrt{1+tan^2(\theta_W)} $, for the derived and
 general values of $g$ and $g^\prime$. The additional neutral orthogonal vector and pseudo-vector
combination is $Z_\mu^\prime =[\sqrt{3/2}( L+2  Q
)-\frac{1}{\sqrt{2}}Z_0]\gamma_0\gamma_\mu$,
 and its expectation value $M_F$ in Eq. \ref{ratioM} gives
 $M_{Z^\prime}=\sqrt{2} M_Z$.  $M_{Z^\prime}$ is  unstable under changes of some of the $Z_\mu^\prime$
 components\footnote{$M_F$ is not a constant or minimum upon
 variation of these parameters.} ($Z_\mu$, $W_\mu$ are stable) which casts doubt on whether it represents a
 physical particle.
 \renewcommand{\arraystretch}{1.5}
\begin{table}[h]
\center{ \begin{tabular}{|c| c c c c c c |} \hline
   & {$fer$}  & $\nu\ \ \ $ & $A_\mu$ & $W^\pm_\mu$   & $Z_\mu$ & $H$  \\
\hline $\{tr [ \tilde H,F]^\dagger_\pm [\tilde H,F]_\pm\}^{1/2}$ &
$M_W    $  & $0   $ & $0$ & $M_W $ & $\sqrt{4/3}M_W $ &
$\sqrt{2}M_W $  \\  $[$ GeV $]$  & $ 80.4$& $0$ &$0$ & $80.4$ &
$92.8$& $113.7$\\ \hline SM   & $-$  & $0$ &$0$ & $M_W$ &
$\sqrt{1+tan^2(\theta_W) }M_W$& $\frac{2\sqrt{2\lambda}}{g}M_W$\\
$[$ GeV $]$
    & $-$   & $0$   & $0$ & $80.4$ & $91.2$& $-$\\ \hline
\end{tabular}}
\caption{\label{tab:tablejbn} Predictions from $M_F$ in Eq.
\ref{ratioM}, based on  the $sin^2(\theta_W)=.25$ result, for
masses of massive fermion ($fer$), neutrino ($\nu$), photon
($A_\mu),$ $W$, $Z$, and Higgs ($H$) particles, in terms of $M_W$,
as compared to the SM, with numerical values for both cases.}
\end{table}
\renewcommand{\arraystretch}{1}
\normalsize
\baselineskip 22pt \vfil\eject \noindent

Application of Eq. \ref{ratioM} to the state with scalar
polarization  $H= \frac{g}{\sqrt{2}}(n_0+\tilde
n_0)=-\frac{1}{2}M_3\gamma_0$  leads to $[\tilde H,H]=\frac{g
v^\prime }{2}(-1+i)[ \tilde n_0, n_0]=\frac{g v^\prime}
{64}(-1+i)[\gamma_0I_+,\gamma_0I_-]=i\frac{g v^\prime
}{8}(-1+i)(1-I)\gamma_5 $, implying $M_H=\sqrt{2}M_W$. The choice
 $\tilde H$ in Eq. \ref{ratioMvscale}    gives mass to  the
hermitian scalar combinations $M_i \gamma_0$, $i=3,4$ in Eq.
\ref{geneqIII}, and
 is stable
 ($n_0\pm  \bar n_0$ makes them either massless or unstable).
  It is pertinent to compare this value with recent
experiments that indicate a Higgs particle exists with a 114-114.5
GeV mass\cite{higgs}.

It is illustrative to derive the same results   within the SM, by
using its equivalence to a formulation in terms of the fields'
underlying spin structure. We consider the Lagrangian component
that gives masses to the electroweak vector fields after
spontaneous symmetry breaking\cite{Weinberg}:
\begin{eqnarray}
\label{Lagramass}
 { \mathcal L}_{M}&=&{\bar v}^\dagger (i  \frac {g^\prime}{2}\bar B^\mu+i\frac {g}{2}\tau^i
\bar  W^{\mu i} )^\dagger(i  \frac {g^\prime}{2}\bar B_\mu+i \frac
{g}{2}\tau^i \bar  W_\mu^i)\bar v \\ \nonumber &=& [...]^\dagger[i
e \bar A_\mu\frac {1}{2}(1+\tau^3)+ \frac{i}{2\sqrt{{g}^2
+{g^\prime}^2}} \bar Z_\mu({g^\prime}^2-g^2\tau^3) +i\frac
{g}{2\sqrt{2}} \bar W^{+}_\mu\tau^+ ]\bar v,
\end{eqnarray}
where $\tau^i$ are the Pauli matrices, we use Eq.  \ref{photon},
$\bar Z_\mu=\frac{1}{\sqrt{g^2+g^{\prime 2}}}(g^\prime \bar
B_\mu-g \bar W^0_\mu)$, $\bar W^\pm_\mu=\frac{1}{\sqrt{2}}(\bar
W_\mu^1\mp i \bar W_\mu^2)$, $\tau^\pm= \tau^1\pm i \tau^2 $,
$e=\frac{g g^\prime }{\sqrt{g^2+g^{\prime 2}}}$ is  the electric
charge,  $\bar v= \frac{1}{\sqrt{2}}\left(
\begin{array}{cc}
 0\\
  v
\end{array} \right)$,  and $v$ the Higgs vacuum expectation value.
$g$ and $g^\prime$ can again either represent coupling parameters
or specific numbers linked to the polarization normalization.
${\mathcal L}_{M}$ can be equivalently written  in terms of the
5+1 $d$ generators
\begin{eqnarray}
\label{Lagramassspin}
 { \mathcal L}_{M}&=&
\frac{v^2}{2}tr  [...]^\dagger[i e \bar A_\mu(I_3+\frac {1}{2}Y)+
\frac{i}{\sqrt{{g}^2 +{g^\prime}^2}} \bar Z_\mu({g^\prime}^2
\frac{1}{2}Y-g^2 I_3) + \\ \nonumber & & i\frac {g}{\sqrt{2}} \bar
W^{+}_\mu I_+  ]\gamma_0\gamma^\mu \tilde H
\\ \nonumber & = &\frac{1}{2} tr  (...)^\dagger( \bar Z_\mu [\tilde H, Z^\mu]_\pm + \sqrt{2} \bar W^+_\mu
[\tilde H, { W}^{\mu +}]_\pm ),
\end{eqnarray}
In the last equality we use a commutator or anticommutator for
$\mu=0$ or $\mu=1,2,3$, respectively. Their expectation values
give the $M_Z$ and $M_W$ masses as in Eq. \ref{ratioM}. The SM
sets $v^\prime=v$ in $\tilde  H$. Thus, the vector-boson masses
can also be interpreted as stemming from a single mass term
defined on a spin space.

 On the other hand, the
Higgs $\bar \rho$ Lagrangian can be similarly written
\begin{eqnarray}
\label{LagraHiggs}
 { \mathcal
 L}_{H}=\frac{1}{2}\partial_\mu\bar \rho\partial^\mu\bar\rho-V\left
 [\frac{(\bar\rho+v)^2}{2}\right],
\end{eqnarray}
with $V[{\bar\phi}^\dagger \bar \phi]=\mu^2 \bar\phi^\dagger \bar
\phi+\lambda (\bar\phi^\dagger \bar\phi)^2$ the potential of the
Higgs doublet, $\mu^2< 0.$ In particular, after spontaneous
symmetry breaking, the mass term can be written
\begin{eqnarray}
\label{LagraHiggsmass}
 { \mathcal L}_{M_H}=-\frac{1}{2} 2 \mu^2\bar\rho^2=-\frac{1}{2}  tr[...]^\dagger[\tilde H,
 H]\bar\rho^2,
\end{eqnarray}
with $v^2=-\mu^2/\lambda$. The second equality has been set so
that ${\mathcal L}_{M}$ in Eq. \ref{Lagramass}   and ${ \mathcal
 L}_{M_H}$ in Eq. \ref{LagraHiggsmass}
exhibit the same underlying operator, with   a symmetry  assumed
under the exchange of fields occupying the same (spin) space.
 Thus, a  unique term is understood to
produce masses at tree level, both among the vector and the scalar
fields, setting the Higgs  parameter $\mu$.

\section{ Vector  fermion-current vertices}
 The  spin 1/2
components form a charge $q=1$ (from $Q$) massive  Dirac particle
$l$ (left-handed part $l_L$)
 and a $q=0$ particle $\nu$,  which remains  massless.
This is as occurs in the SM for charged leptons   and neutrinos,
to be identified respectively with $l^-$ and $\nu$. Assuming the
same spin symmetry for the fermion-mass term  implies taking $H
l^-=\frac{1}{\sqrt{2}}[n_0+\tilde n_0,l^-]= \frac{1}{2}l^-$,
where, {\it e.g.}, $l^-=\frac{1}{\sqrt{2}}(l_R^-+l_L^-)$, $l_R^-$
is given in Eq. \ref{lR}, $l_L^-= \frac{1}{8} (1-I \gamma _5 ) (
1+I) (\gamma_1+ i I \gamma_2)$, with equivalent results with
$\tilde H$ in $tr[...]^\dagger [\tilde H,l^-]$. Although an $M_W$
mass is obtained for the massive leptons, extended
 models accounting for quarks\cite{Jaimenext} will presumably give a similar prediction,
  so that a connection to the top-quark mass scale may be at hand.

 SM
vertices  describing the fermions to the vector-fermion couplings
are obtained from the argument  after Eq. \ref{inteLagratra}.
Following it, the  interaction of the $l$, $\nu$, and $W^\pm$
particles can be described through the Lagrangian
\begin{eqnarray}
\label{Lagrangianferver} {\mathcal{L}}_{Wint}&=&\bar W^+_{\mu}
\frac{g}{2\sqrt{2}}[\bar \nu^\dagger(1-
\gamma_5)\gamma_0\gamma^\mu \bar l_L^-+hc]\\ \nonumber &=&\bar
W^+_{\mu} tr\frac{g}{2\sqrt{2}}[\nu^\dagger(1-I
\gamma_5)\gamma_0\gamma^\mu l_L^-+hc] \\ \nonumber &=& \bar
W^{+\mu} tr[\nu^\dagger W^+_\mu l_L^-+hc],
\end{eqnarray}
where in the second equality  5+1 spin polarizations matrices  are
used, as in  Eq. \ref{inteLagratra}.
  Similar expressions as in the SM can be
obtained for the lepton interactions with the $Z_\mu$ particle.

 In addition,  the vertices  give information on the coupling constants  $g^\prime$ and
 $g$,
 which can also be extracted from the normalization constants of
 the vector bosons  in Section 5.
 $g$ can be obtained from
the coupling of   the massive charged vectors  $W_{\mu}^+$ and the
charged current, defined by the neutrino and the charged leptons
wave functions, represented in Eq. \ref{Lagrangianferver}. These
are $g=1,$    $g^\prime=1/\sqrt{3}$. The next Clifford algebra
model at 7+1 $d$, reproduces the same results and, under the
assumption that $K=1$, one finds a $ \frac{1}{\sqrt{2}}$ factor in
the coupling obtained from the trace in
 Eq. \ref{pointproduct}, with respect to the lower-dimensional algebra, with
 predictions
 $g=1/\sqrt{2}\approx .707$  $g^\prime=1/\sqrt{6}\approx
.408$.
It is a consistency feature of the theory that any these values
are in accordance with $\theta_W$.
 In addition, these values are to be compared with the experimentally measured ones
at energies  of the mass of the $Z_\mu$ particle, which is where
the breakdown  of the $SU(2)_L\times U(1)_Y$ symmetry occurs.
These are $g_{exp}^\prime\approx .36$, $g_{exp}\approx .65$, and
$sin^2(\theta_{Wexp})\approx .23$\ . Closer numbers are obtained
when considering the SM strong sector\cite{Jaimenext}.

 The theory thus presented succeeds in
reproducing many aspects of the electroweak sector of the standard
model. We conclude that the latter allows for a consistent spin
interpretation  that  enriches it with relevant information on the
couplings, the representations, and the masses. The close
connection between the results hence derived and the physical
particles' phenomenology makes plausible the idea that, as with
the spin,
  the gauge vector and matter fields, and their
interactions  originate in the structure of space-time.

{\bf Acknowledegment}

 The
author acknowledges support from  DGAPA-UNAM through project
IN118600.

\setcounter{equation}{0}







\end{document}